\documentclass[11pt,a4paper]{article}

\usepackage{authblk}
\usepackage[english]{babel}
\usepackage{amsmath}
\usepackage{graphicx}
\usepackage[colorlinks=true, allcolors=blue]{hyperref}
\usepackage{xr}
\usepackage{subcaption}
\usepackage[margin=2.2cm]{geometry}

\usepackage{amssymb}
\usepackage{float}
\usepackage{siunitx}
\usepackage{multirow}
\usepackage{subcaption}
\usepackage{amsmath}
\usepackage{orcidlink}
\usepackage[numbers, sort&compress]{natbib}
\DeclareSIUnit{\atom}{atom}
\DeclareSIUnit\angstrom{\text {Å}}
\title{Comparing fine-tuning strategies for machine learning force fields in lithium-ion diffusion}

\author[1,*]{Nada Alghamdi\orcidlink{0000-0001-5638-3483}}
\author[1]{Paolo de Angelis\orcidlink{0000-0003-1866-2988}}
\author[1,2]{Pietro Asinari\orcidlink{0000-0003-1814-3846}}
\author[1,2,*]{Eliodoro Chiavazzo\orcidlink{0000-0001-6165-7434}}

\affil[1]{Department of Energy, Politecnico di Torino, Corso Duca Degli Abruzzi, 10129 Torino, Italy}
\affil[2]{Istituto Nazionale di Ricerca Metrologica, Strada delle Cacce 91, 10135 Torino, Italy}

\date{}

\begin{document}
\maketitle
\vspace{-2.5em}
\begin{center}
*Corresponding authors: \texttt{nada.alghamdi@polito.it}, \texttt{eliodoro.chiavazzo@polito.it}
\end{center}

\begin{abstract}
Machine learning force fields (MLFFs) are transforming materials science and engineering by enabling the study of complex phenomena, such as those critical to battery operation.
In this work, we explore the predictive capabilities of pre-trained and fine-tuned MACE MLFF and compare different fine-tuning strategies for predicting interstitial lithium diffusivity in LiF, a key component in the solid electrolyte interphase in Li-ion batteries.
Our results demonstrate that the MACE-MPA-0 foundational model achieves comparable accuracy to well-trained DeePMD,  in predicting key diffusion properties based on large scale molecular dynamics simulation, while requiring minimal or no training data.
For instance, the MACE-MPA-0 predicts an activation energy $E_a$ of \SI{0.22}{\electronvolt}, the fine-tuned model with only 300 data points predicts  $E_a = $ \SI{0.20}{\electronvolt}, both of which show good agreement with the DeePMD model reference value of $E_a = $ \SI{0.24}{\electronvolt}.
In this work, we provide a solid test case where fine-tuning approaches, whether using data generated for DeePMD or data produced by the foundational MACE model itself, yield similar robust performance to the DeePMD potential trained with over 40,000 actively learned data, albeit requiring only a fraction of the training data.
\end{abstract}

\section{Introduction}
High-performance energy-storage systems are key technologies for the energy transition. 
In particular, Lithium-ion batteries (LIBs) are recognized to play a crucial role in renewable energy storage, grid stability, and low-carbon mobility~\cite{balakrishnan_great_2021, amici_roadmap_2022, yoshio_lithium-ion_2009}.
The performance, lifetime, and safety of LIB cells are governed by processes at the electrode–electrolyte interface, where the solid-electrolyte interphase (SEI) forms~\cite{goodenough_challenges_2010, shan_brief_2021}. 
The SEI is a reactive passivation layer where multiple chemical reactions occur.
It is highly heterogeneous, comprising many organic and inorganic components.
The SEI extends the electrolyte's electrochemical window and isolates the anode, enabling proper operation of the LIB~\cite{goodenough_challenges_2010}.
However, uncontrolled formation or growth can degrade cell performance, leading to capacity fade and, in extreme cases, thermal runaway~\cite{wang_fluorination_2024}.
Recent experimental evidence showed that LiF-rich SEIs enhance mechanical stability, provide electronic insulation, and reduce interfacial resistance, which translates into improved performance and safety for fluoride-rich battery formulations \cite{wang_fluorination_2024}. 
Moreover, controlling SEI formation is crucial not only for Li-ion batteries but also for beyond-Li-ion systems, such as Ca-ion batteries~\cite{Palacin_2024, Titirici_2024}.
Therefore, it is important to understand and engineer the SEI for improved battery lifetime, safety, and performance~\cite{cappabianca2023overview, deAngelis2024enhancing}.

The advent of Machine Learning Force Fields (MLFFs) or synonymically Machine Learning Interatomic Potentials (MLIPs) has significantly advanced the field of materials modeling~\cite{Generalized2007Behler, Unke2021Machine, Deringer_2020}, enabling simulations over longer timescales and larger systems that were previously unaffordable.
Among many applications, it plays a crucial role in reactive atomistic simulations that are essential for studying physical phenomena of relevance for energy conversion processes \cite{CASINI2024100330}.
MLIPs offer a promising solution for addressing complex simulations, where understanding the interplay between components (e.g. the electrode and electrolyte, the SEI) is crucial to ensure battery stability and safety, but it remains inaccessible with traditional experimental methods~\cite{amici_roadmap_2022, cappabianca2023overview}, and computationally expensive to perform with traditional {\it ab-initio} methods~\cite{battery2030}.
Many MLIPs have emerged in recent years. 
Their development can be viewed as a progressive refinement in how atomic environment description (descriptors) and regression model architecture are constructed, or in other words, how symmetry and many-body correlations are encoded into the models .
The introduction of Behler-Parrinello Neural Network (BPNN)~\cite{Generalized2007Behler} marked a foundational contribution in the field. 
It introduced two key ideas that remain central to ongoing development: 
First, decomposing the total energy $E$ into a sum of atomic contributions $E = \sum_i E_i$.
This allows for simulating systems of arbitrary size with linear scaling and ensures permutation invariance by sharing the Neural networks (NNs) for atoms of the same chemical species. 
Second, introducing descriptors that explicitly encode the required physical invariances (translational and rotational),  which are manually engineered~\cite{Generalized2007Behler}.
Building on that, the Deep Potential Molecular Dynamics (DeePMD) model~\cite{Zhang2018Deep, deepmd} used neural networks for both regression and descriptors, i.e., instead of using fixed, manually engineered descriptors, both the representations and regression model are now learned directly from the data and optimized during training.
In this approach, the atomic environment for each atom $i$ is represented as a transformation of the coordinates from the global configuration to a local reference frame centered around the atom. 
Accordingly, a mapping from the atomic positions, bond angles and chemical species, to local vector descriptors $\boldsymbol{D}_i$ is learned while preserving translation, rotational, and permutational invariance~\cite{Zeng2023DeePMD}.
Similar to the BPNN, in DeePMD, the total potential energy $E_{tot}$ of a given atomic configuration is the sum of the energy contribution $E_i$ of each atom $i$ in the system.
For a system of several atomic species, each atomic species $s$, has its own neural network NN$_s$, and every atom $i_s$ within that species uses the same NN$_s$ of the species, so the total energy:
\begin{equation}
    E_{tot} = \sum_{i=1}^{N_{atom}} E_i
    \quad \Longleftrightarrow \quad	 E_{tot} = \sum_{s=1}^{N_{species}} 
    \sum_{i_s}^{N_{atom\in s}} 
    E_{NN_s}(\boldsymbol{D}_{i_s}),
\end{equation}
where $N_{atom}$ is the total number of atoms in the system, $N_{species}$ is the total number of atomic species, $N_{atom\in s}$ is the number of atoms belonging to species $s$, and $E_{NN_s}(\boldsymbol{D}_{i_s})$ is the energy predicted by the NN$_s$ associated with species $s$ for the atom with descriptor $\boldsymbol{D}_{i_s}$ \cite{Wang_2018, Thiemann_2025}.
Subsequently, models based on Graph Neural Network (GNN), particularly Message-Passing Neural Network (MPNN)~\cite{gilmer2017neuralmessagepassingquantum}, have emerged as a framework for MLFFs. 
Among them MACE~\cite{batatia2023macehigherorderequivariant} which combines the Atomic Cluster Expansion (ACE)~\cite{ACE}, that forms a complete representation of local environment, with an equivariant MPNN, that allows for message passing hence the atoms exchange information with their neighbors and thus learn hierarchical many-body interactions.
In principle, this approach promises improved data efficiency, as the relevant physical symmetries are embedded directly in the model architecture rather than being inferred from the training data, combined with improved learning of many-body interactions~\cite{batatia2023macehigherorderequivariant}.
This is in contrast to the DeePMD, where learning many-body interaction happens only implicitly within a single learned invariant descriptors.

More specifically, in the MACE model a material is represented as a graph in which each node corresponds to an atom, and each edge represents a connection between atoms within a specific cutoff radius. 
The features are built using learnable radial functions, spherical harmonics, and learnable embedding of neighboring node features. 
Equivariance is preserved through symmetrization with Clebsch–Gordan coefficients.
Subsequently, the higher-order features are generated by a tensor product of these equivariance features. 
Symmetrizing by generalized  Clebsch-Gordan coefficients maintains correct equivariance and allows for very efficient computation of higher order features as these coefficients are highly sparse and can be pre-computed.
This constitutes the body-ordered basis for message construction, where incorporating higher body-order features in the messages allows the model to capture higher correlation order with fewer message-passing steps~\cite{batatia2023macehigherorderequivariant}.
In addition, the graph representation and the tensor product decomposition lends itself to an efficient implementation on GPU architectures, making the approach highly scalable.

By pre-training MACE on large chemical spaces, emerging foundational models encapsulate broad knowledge of atomic interactions.
And they demonstrated notable success in describing different chemistries~\cite{batatia2024foundationmodelatomisticmaterials, Haochen2024systematic, Bruno2025performance}.
Moreover, these foundaitonal models can later be customized for specific tasks or used directly for fast though less accurate simulations.
In spirit, this is similar to large language models (LLMs), trained on the entire internet, that serves as general purpose models.
MACE model, MACE-MP-0, trained on the Materials Project trajectory dataset (MPtrj) \cite{Deng2023CHGNet}, containing over 1.5 million data point, exhibited excellent extrapolation capability to out of distribution domains \cite{batatia2024foundationmodelatomisticmaterials}. 
Several foundational MACE models trained on different datasets have followed,
MACE-MPA-0, which includes 3.5 million new configurations, combines MPtrj with a subset of the Alexandria dataset \cite{Alex}. 
MACE-OMAT-0, trained using Open Materials 2024 (OMat24) dataset \cite{barrosoluque2024openmaterials2024omat24}, containing over 110 million Density Functional Theory (DFT) single-point calculations of inorganic materials, including a wide range of non-equilibrium structures, other foundational models based on different training datasets are continually emerging.

Similar to recent approaches of transfer learning in LLMs and computer vision models \cite{Church_Chen_Ma_2021}, fine-tuning foundational models has emerged as an efficient strategy for achieving high accuracy while substantially reducing training data requirements.
By leveraging the broad knowledge encoded in a foundation model (pre-trained on extensive datasets spanning diverse materials), fine-tuning enables rapid specialization to specific sub-domains with minimal additional data \cite{D4FD00107A}.

Until recently, atomistic modeling of the SEI formation was challenging due to the reactive and chemically complex nature of the process~\cite{BHOWMIK2019446, WU2021106489, deAngelis2024enhancing, Zhang_2023}.
Classical force fields, which rely on fixed topology, are inherently unable to describe bond breaking and formation, limiting their applicability to reactive environments~\cite{senftle2016reaxff}.
Reactive force fields such as ReaxFF~\cite{vanDuin2001ReaxFF,senftle2016reaxff} represented an important step forward, however, their development requires substantial parametrization effort, and their predictive accuracy remains system dependent~\cite{deAngelis2024enhancing, senftle2016reaxff}.
On the other hand, {\it Ab Inito} Molecular Dynamics (AIMD) simulations, are highly accurate but computationally demanding, hence, they are typically restricted to systems containing at most a few hundred atoms and to simulation times on the order of picoseconds~\cite{senftle2016reaxff, li2025enabling}.
The emergence of MLFF, has significantly expanded the scope of tractable problems~\cite{Deringer_2020}.
Recently,  Li {\it et al.}~\cite{li2025enabling}, studied lithium ion diffusion mechanisms in Li$_2$CO$_3$ and Li$_2$EDC, two important SEI components, using the Moment Tensor Potential (MTP), using several iterations of active learning to construct a training set of 8,686 structures.
They find diffusivity and activation energy values in good agreement with the reference.
Similarly, De Angelis {\it et al.}~\cite{deangelis2025} employed a DeePMD potential to study lithium diffusivity in LiF, following an active learning protocol that resulted in a dataset of approximately 40,000 configurations.
They capture interstitial and vacancy diffusion processes, and suggest a ring diffusion mechanism at  high concentration~\cite{deangelis2025}.
These studies of isolated SEI components can provide valuable insights, however, extending simulations to capture a full SEI formation process will require, in addition to accurate and computationally efficient models, an improved data efficiency in the training of MLFFs.

In this study, we consider the MACE foundational model and its fine-tuned variants, motivated by the reported good out-of-the-box performance~\cite{batatia2024foundationmodelatomisticmaterials} and expected data efficiency~\cite{kovacs2023evaluation}, to predict lithium diffusivity in LiF as a representative model system for SEI-related processes.
We consider the foundational model MACE-MPA-0 and several fine-tuned variants, covering both the reuse of data generated during the active learning of a DeePMD model~\cite{deangelis2025} and the construction of a dataset via an active learning protocol tailored for this model.
Although energy and force regression errors are informative indicators of fitting quality, however, they do not necessarily guarantee reliable performance under realistic finite-temperature dynamical conditions~\cite{fu2023forcesenoughbenchmarkcritical}.
More recent efforts reflect a broader shift toward evaluation frameworks that extend to physically relevant performance metrics~\cite{riebesell2025framework, Yu2024Systematic, Wines2025CHIPS, Li2025Are, fu2023forcesenoughbenchmarkcritical, chiang2025mliparenaadvancingfairness, Bruno2025performance}.
However, these studies mainly consider pretrained foundational models, without providing systematic analysis of fine-tuning strategies and without emphasizing long MD simulations.
While Focassio~{\it et al}~\cite{Bruno2025performance} assessed foundational MACE and M3GNet models, including a fine-tuned variant of them, their evaluation is largely limited to surface energy calculations, and the impact of dataset size and composition was not analyzed.
On the other hand, our benchmarking strategy assess the model through its ability to reproduce a physically meaningful process obtained from long MD simulations extending from 3 to 9~ns. Such timescales allow us to probe transport properties and long-term stability of the potential.
This provides a more stringent and physically relevant validation and is able to reveal accumulated errors or stability issues over extended trajectories.
Moreover, we systematically compare different fine-tuning strategies and explicitly address reusability of data and computational efficiency during fine-tuning.
This aspect is rarely considered in existing approaches, yet it is a crucial aspect for investigating complex, multi-component phenomena such as SEI formation, which involves multiple species including LiF, Li$_2$O and Li$_2$CO$_3$ among many others.

\section{Method}
Molecular Dynamics (MD) simulations were performed using the Large-scale Atomic/Molecular Massively Parallel Simulator (LAMMPS)~\cite{LAMMPS}, the Atomic Simulation Environment (ASE)~\cite{ase-paper}, and MACE machine learning force fields~\cite{batatia2023macehigherorderequivariant, Batatia2022Design}. 
Simulations were carried out in the NVT ensemble with a time step of 1~fs.
A 5×5×5 supercell of the LiF conventional bulk (containing 1000 atoms and measuring 20.42 \AA~in length) was employed for the diffusion calculations.
Temperature was controlled by the Nosé–Hoover thermostat \cite{nose1984unified, hoover1985canonical} with a time constant of $\SI{0.1}{\pico\second}$.
The total simulation time for the activation energy calculations was $\SI{9}{\nano\second}$ using four replicas for each temperature. 
Furthermore, to compare models performance, diffusivity was calculated at 400 K and 450 K for 3 ns, with two independent replicas.
An equilibration period of 300 ps was used for the MACE-MPA-0 trajectory. 
For all other simulations, a 100 ps equilibration time was employed, the starting point, however, was from the equilibrated structures obtained from the MACE-MPA-0 simulation.
This equilibration time ensures that the system reaches stable fluctuations in both energy and temperature (see Section S3 of the Supplementary Information for more details).
As a preliminary assessment, we evaluated the bulk equilibrium properties. 
We find that the bulk equilibrium properties are accurately predicted by MACE (details are provided in Table. S1 in the Supplementary Information), consistent with its training on periodic bulk structures from the Materials Project, which covers many Li and F-containing systems \cite{batatia2024foundationmodelatomisticmaterials, Deng2023CHGNet}.
All input files and training data required to reproduce the results are publicly available (see Data Availability section).
\subsection{Diffusivity}
The diffusivity was found according to the Einstein-Smoluchowski relation:

\begin{equation}
\lim_{t \rightarrow \infty}\left\langle\left\|\mathbf{r}(t)-\mathbf{r}(0)\right\|^2\right\rangle=c_d D t
\end{equation}

where \( t \) is time, \( D \) is the diffusion coefficient, and \( c_d \) is a constant that accounts for the system dimensionality; being \( c_d = 6 \) for three-dimensional diffusion.
For a sufficiently long simulation time, diffusion coefficients are extracted by calculating the slope of the Mean Square Displacement (MSD) versus time for each simulation.
The activation energy for diffusion, \( E_a \), was then obtained by fitting the temperature dependence of \( D \) to the Arrhenius like equation:

\begin{equation}
D = D_0 \cdot \exp \left( -\frac{E_a}{k_B T} \right),
\label{Arrhenius}
\end{equation}

where \( D_0 \) is the maximum diffusivity, \( k_B \) is the Boltzmann constant, and \( T \) is the temperature~\cite{allen2017computer}. 
Note that finding the pre-exponential factor is not trivial by heuristic arguments \cite{Mehrer2007}. Here we find it through an empirical measurement of {\it in silico} MD experiments, this has an advantage over other methods as it inherently accounts for entropic effects.

\subsection{Fine-tuning with DeePMD data}
\label{Fine-tuning with DeePMD data}
The MACE-MPA-0 model \cite{mace_mpa_0} served as the starting point for our experiments.
First, we fine-tune it using data generated during the active learning process of DeePMD (For details about the dataset creation and access, see Ref.~\cite{deangelis2025}.)
The dataset contains over 40{,}000 structures, from which we randomly select a small subset for fine-tuning.
We select fint-tuning datasets of increasing size: 110, 210, 410, 710, 800 and 1600 samples. 
Additionally, we incorporate varying numbers of Materials Project data—0 (naive fine-tuning), and multi-head fine-tuning with 100, 1000, and 10,000 data points—into the training.
We explore different numbers and ratios of bulk, Li interstitial and MPtrj pre-training datasets.
Details of the composition of the different training datasets considered are shown in Table~\ref{tab:finetuning_setup}.
Notably, the 800 sample case has identical structures to the 710 sample case, it only includes 90 additional bulk structures.
\begingroup
\renewcommand{\arraystretch}{1.1} 
\begin{table}[htp!]
\centering
\caption{Summary of datasets used to fine-tune MACE-MPA-0 \cite{mace_mpa_0}, for the case where we used configurations from DeePMD actively learned dataset~\cite{deangelis2025} to fine-tune with a varying amount from the pre-training Materials Project (MPtrj) \cite{Deng2023CHGNet} dataset.}
\begin{tabular}{lcc|c}
\hline 
\multicolumn{3}{c|}{Fine-tuning data~\cite{deangelis2025}  } & Pre-training data~\cite{Deng2023CHGNet}
\\\hline
 Bulk &  Interstitial & Total & MPtrj \\
LiF & Li in LiF  & & \\
\hline 
10  & 100 & 110 & 1000 \\
10  & 200 & 210 & 0, 100, 1000\\
10  & 400 & 410 & 1000\\
10  & 700 & 710  & 0, 100, 1000, 10000\\
100 & 700 & 800  &  0, 100, 1000, 10000 \\
200 & 1400 & 1600 & 1000 \\
\hline
\end{tabular}
\label{tab:finetuning_setup}
\end{table}
\endgroup

Training was done for 800 epochs; however, the 10,000 pre-training cases, being significantly larger, was limited to 200 epochs. 
A summary is reported in Table. \ref{tab:finetuning_setup}, while the corresponding learning curves are provided in the Supplementary Information.

The weighted sum of Huber loss~\cite{Huber1992} functions for energies, forces, and stresses were used as a training loss. The general form of the Huber loss for a difference $x_n-y_n$ is given by:
\begin{equation}
\mathcal{L}_\delta(x_n-y_n)= \begin{cases}\frac{1}{2} (x_n-y_n)^2, & \text { for }|x_n-y_n| \leq \delta, \\ \delta \cdot\left(|x_n-y_n|-\frac{1}{2} \delta\right), & \text { otherwise } .\end{cases}
\end{equation}
The respective differences $x_n-y_n$ are the error between model predictions and reference values for energies, forces and stresses.
Following the universal loss in MACE \cite{batatia2024foundationmodelatomisticmaterials}, Huber deltas $\delta_E,\ \delta_{F_0}, \ \delta_{\sigma}$ are set to 0.01,
and conditional Huber loss was used for the forces such that the Huber delta $\delta_{F}$ adapts based on the magnitude of the force on each atom. The Huber delta decreases as the force component magnitude increases according to:
\begin{equation}
\delta_{F_\alpha} =
\begin{cases}
\delta_{F_0}, & F_\alpha < \SI{100}{(\electronvolt\per\angstrom\per\atom)} \\[6pt]
0.7\,\delta_{F_0}, & \SI{100}{} \leq F_\alpha < \SI{200}{(\electronvolt\per\angstrom\per\atom)} \\[6pt]
0.4\,\delta_{F_0}, & \SI{200}{} \leq F_\alpha < \SI{300}{(\electronvolt\per\angstrom\per\atom)} \\[6pt]
0.1\,\delta_{F_0}, & F_\alpha \geq \SI{300}{(\electronvolt\per\angstrom\per\atom)}
\end{cases}
\end{equation}

where $F_\alpha$ represents the force component in each Cartesian direction ($\alpha = x, y, z$) \cite{batatia2023macehigherorderequivariant, batatia2024foundationmodelatomisticmaterials, Batatia2022Design}.
The weights for energy, forces, and stresses losses were set to $(\lambda_E, \lambda_F, \lambda_\sigma) = (1, 10, 1)$, assigning higher importance to forces as they were harder to learn.


\subsection{Active learning}
\label{sec:active_learning}
An active learning protocol similar to \cite{deangelis2025, yang2023reactant} was followed.
A committee of four models were trained using the same data, with different random initialization, to quantify model uncertainty and inform decisions regarding additional training data.
Simulations were performed with an NVT ensemble, consistent with previous MD simulations, using 3×3×3 supercell (comprising 216 atoms, 12.25 \AA).
Each simulation was carried out for 24 hours, corresponding to approximately 600 ps trajectories.
The mean force error was evaluated as \cite{Zeng2023DeePMD}:
\begin{equation}
\sigma=\max _i\left[\sqrt{\frac{1}{N} \sum_{k=1}^N\left\|\mathbf{F}_i^k-\overline{\mathbf{F}}_i\right\|^2}\right],
\end{equation}

where the size of the committee $N=4$, $\mathbf{F}_i^k$ is the force on $i$-th atom predicted by the $k$-th model, and $\overline{\mathbf{F}}_i$ is the committee average \cite{Zeng2023DeePMD, deangelis2025}.
The structures were sampled according to the uncertainty: 30\% of the data are taken from within the 1$\sigma$ range, 60\% within the 1.5$\sigma$ range, and the remaining within the 2$\sigma$ range \cite{yang2023reactant}. 


\subsection{Fine-tuning with MACE-MPA-0 data}
\label{MPA-data}
We train two additional models referred to as Fine-Tuned 1 (FT1) and Fine-Tuned 2 (FT2), starting from data sampled from configurations explored by the MACE-MPA-0 trajectories, using naive fine-tuning without incorporating any pre-training data.
Energy, forces and stresses for new sampled configurations were evaluated with DFT using the Quantum ESPRESSO code \cite{giannozzi2009quantum, Giannozzi_2017, Giannozzi2020exascale}.
We used the PBE exchange-correlation functional and 
Projector Augmented Wave (PAW) pseudopotentials \cite{PhysRevLett.77.3865}.
Brillouin-zone sampling was limited to the $\Gamma$-point. 
Self-consistent convergence threshold was set to $5\ \times \ 10^{-7}$ Ry.
Plane-wave basis with a kinetic energy cutoff of 110 Ry and charge-density cutoff of 440 Ry, were used. 
Electronic occupations were treated with Gaussian smearing of 0.005 Ry.
This is consistent with the methodology used to generate the DeePMD dataset~\cite{deangelis2025}.

A summary of the training data is shown in Table. \ref{tab:active_setup}.
A total of 76 structures are randomly selected: 16 bulk configurations, 4 structures at each temperature: 300, 400, 500, and 600 $K$, and 60 interstitial configurations, 10 structures at each temperature: 300, 350, 400, 450, 500, and 600 $K$.
For the FT1 model, we perform two active learning steps. 
In each step, 40 additional structures are selected based on committee disagreement in the interstitial Li in the LiF configuration. 
In total, FT1 is trained on 156 structures and validated on equal amount of structures.
For the model FT2, we perform only one active learning step. In this case, the training data includes both interstitial Li (10 structures per temperature) and bulk (7 structures per temperature) configurations from the committee run. 
This results in a total of 144 structures used for training and similar amount for validation. 
The final force disagreement among the ensemble of models was on the order of \num{e-5} eV/\AA \ for FT1 and \num{e-4} eV/\AA \ for FT2 (see Supplementary Information S12).

\begingroup
\renewcommand{\arraystretch}{1.1}

\begin{table}[!ht]
\centering
\caption{Dataset creation through active learning for models FT1 and FT2. The starting dataset for the active learning for both FT1 and FT2 models (step 0) was created by sampling configurations from MACE-MPA-0 MD trajectories, for bulk LiF and interstitial Li defects in LiF across a range of temperatures. FT1 model then added interstitial Li configurations in 2 steps, while in FT2, additional configurations for both bulk and interstial Li cases were added in single step. In each case sampling was done from the fine-tuned models themselves}
\begin{tabular}{cc|cc|cc}
\hline 
step &  Temp (K) & \multicolumn{2}{c|}{FT1} & \multicolumn{2}{c}{FT2} \\
&& \small{Bulk} & \small{Li interstitial} & \small{Bulk} & \small{Li interstitial} \\
&& \small{LiF} & \small{in LiF} & \small{LiF} & \small{in LiF}  \\
\hline 
\multirow{6}{*}{0} & 300 &  4 & 10 & 4 & 10 \\
 & 350 &  - & 10 & - & 10 \\
 & 400 &  4 & 10 & 4 & 10 \\
 & 450 &  - & 10 & - & 10 \\
 & 500 &  4 & 10 & 4 & 10 \\
 & 600 &  4 & 10 & 4 & 10 \\
\hline 
\multirow{4}{*}{1} & 300 &  - & 10 & 7 & 10 \\
 & 400 &  - & 10 & 7 & 10 \\
 & 500 &  - & 10 & 7 & 10 \\
 & 600 &  - & 10 & 7 & 10 \\
\hline
\multirow{4}{*}{2} & 300 &  - & 10 & - & - \\
 & 400 &  - & 10 & - & - \\
 & 500 &  - & 10 & - & - \\
 & 600 &  - & 10 & - & - \\
 \hline 
\multicolumn{2}{c|}{Total} &\multicolumn{2}{c|}{156}& \multicolumn{2}{c}{144} \\
\hline
\end{tabular}
\label{tab:active_setup}
\end{table}
\endgroup

The models were trained for 800 epochs using the Huber loss function. 
For these two cases, we found the optimal balance for learning all three components of the loss function by setting the weights to $(\lambda_E, \lambda_F, \lambda_\sigma) = (1, 1, 10)$ for energy, forces, and stresses, respectively .
\section{Results and Discussion}
\begin{figure*}[!th]
    \centering
    \includegraphics[width=\linewidth]{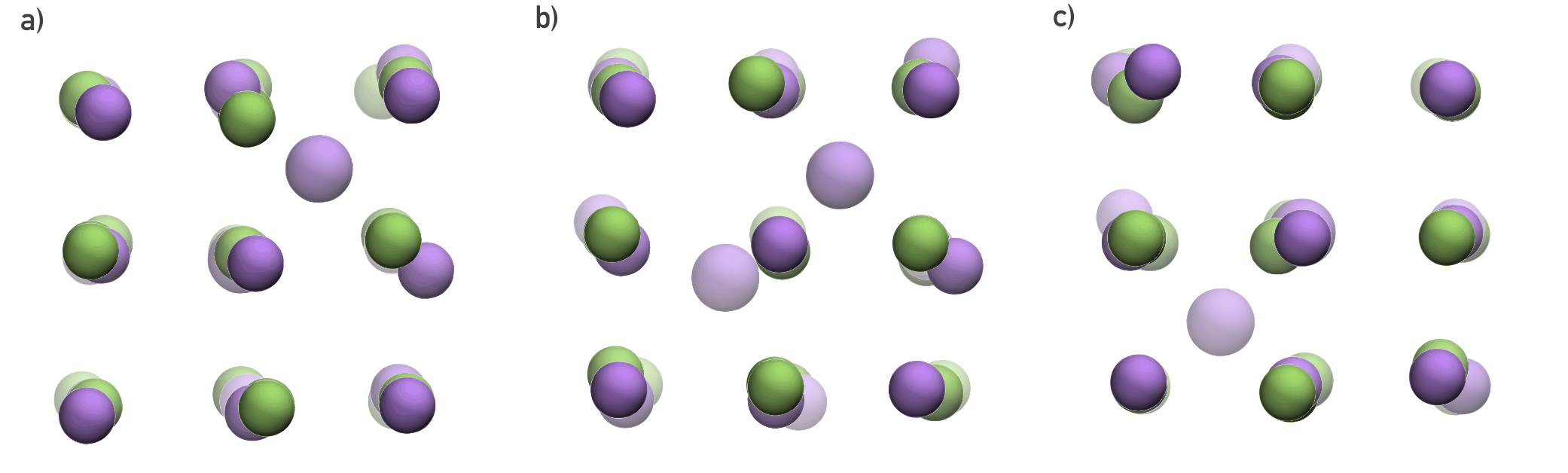}
    \caption{Molecular dynamics snapshots of a knock-off event showing (a) the initial state, (b) an intermediate configuration along the transition pathway, and (c) the final state following the knock-off event.}
    \label{fig:knockoff}
\end{figure*}

As Li-ion propagates through the SEI, several transport mechanisms can take place, including direct interstitial hopping and knock-off hopping, among others~\cite{Tan2021Growing, deangelis2025}.
Herein, we start all MD simulations by placing an interstitial Li ion into bulk LiF (Fig. \ref{fig:knockoff} (a)) without applying any bias.
We let the system evolves to sample the free-energy landscape at finite temperature, hence in our treatment the enthalpic and entropic contributions on transport mechanisms and the accessibility of intermediate states are included.
We find that, in all cases, the system exhibits knock-off hoppings, i.e., an interstitial Li ion knocks a lattice Li atom out of its site, taking its place, while the displaced lattice Li becomes the new interstitial ion, as shown in Fig.~\ref{fig:knockoff}.
In the process, the Li ion breaks bonds with its surrounding atoms and forms new ones.
This is in agreement with studies showing it to be preferable transport mechanism over the direct interstitial hopping~\cite{deangelis2025,zheng2021lithium}.
Achieving sufficient sampling of rare events via MD can be computationally demanding~\cite{he2018statistical}. 
This is precisely where MLIPs provide significant advantage, as they allow access to the long timescales required to observe numerous diffusion events needed for adequate sampling.
In our work, independent diffusion events were observed from simulations spanning 3 to 9 ns with 2 to 4 replicas, providing adequate sampling for the relevant transition pathways.

\subsection{Arrhenius behavior of Li diffusion}
As the transport of Li ions is key to battery performance, accurately describing this process provides an essential test for MLIP applied to battery materials. 
We therefore first benchmark the MACE-MPA-0 and fine-tuned MACE-MPA-0 models by using the Arrhenius equation (eq. \ref{Arrhenius}) to derive their predicted activation energies $E_a$ and maximum diffusivity $D_0$, which we compare against the reference values simulated by the DeePMD model \cite{deangelis2025}.

\begin{figure}[!ht]
    \centering
    \includegraphics[width=0.5\linewidth]{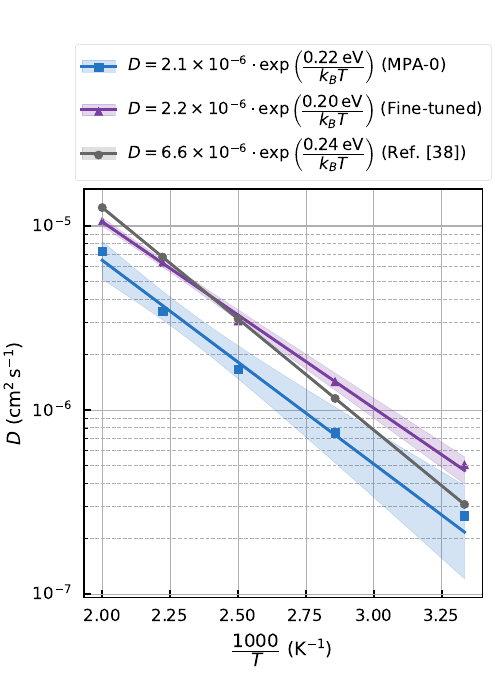}
    \caption{Arrhenius plot for activation energy $E_a$ calculated with MACE-MPA-0 and our fine-tuned model (incorporating 200 DeePMD data points \cite{deangelis2025} and 100 MPtraj pre-training data points, ft200-pt100) from 9 ns MD trajectories. For reference, we include DeePMD results found in Ref. \cite{deangelis2025}.}
    \label{fig: activation energy}
\end{figure}

The Arrhenius plot for the MACE-MPA-0 model and a fine-tuned model along with the reference DeePMD values \cite{deangelis2025} are shown in Fig. \ref{fig: activation energy}.  
The fine-tuned model was trained on 200 data points randomly sampled from actively learned DeePMD structure~\cite{deangelis2025} and 100 pre-training MPtraj~\cite{Deng2023CHGNet} data points, we denote it as ft200-pt100 (ft: fine-tuning, pt: pre-training).
MACE-MPA-0, trained only on equilibrium and near equilibrium data, shows robust out-of-distribution performance in our case of interstitial defects and provides a good prediction of the activation energy $E_a$ of $0.220\ \pm 0.015$ eV.
The Li diffusivity is underestimated at all temperatures. 
The foundational model MACE-OMAT-0, which incorporates a larger portion of non-equilibrium structures, yields values closer to the reference diffusivity (Fig. S7, Supplementary Information).
Notably, fine-tuning with as few as 300 data points results in fair performance, the Arrhenius behavior is maintained with an activation energy value of $0.200\ \pm 0.004$ eV and the underestimation of the diffusivity values is less severe. 
Using approximate Nudged Elastic Band (ApproxNEB) method, Deng {\it et al}~\cite{deng2024overcoming} demonstrated that foundational MLIPs systematically underestimate ion migration barriers compared to DFT with mean absolute error difference of 0.30 eV for MACE-MP-0 using 470 Mg-ion migration pathways.
They observed softening of the Potential Energy Surface (PES) in many foundational MLIP~\cite{deng2024overcoming}.
This effect arises from the underestimation of energies and forces that stems from their training primarily on equilibrium atomic configurations \cite{deng2024overcoming}.
On the other hand, we estimate activation energies by incorporating temperature effects through Arrhenius fitting of diffusivity data.
We find a small underestimation of the activation energies predicted by the MACE-MPA-0 and its fine-tuned variant of just 0.02-0.04 eV relative to the reference.
Table. \ref{tab:comparison} summarizes the performance of MACE-MPA-0 and ft200-pt100 fine-tuned model for activation energy $E_a$, pre-exponential factor $D_0$, and diffusivity $D \ \text{at}\ 450 $ and $500$ K.
\begin{table*}[ht]
\centering
\caption{The predicted values of the activation energy $E_a$, pre-exponential factor $D_0$, and diffusivity $D$ at 450 and 500 $K$, for MACE-MPA-0 foundational model, ft200-pt100 fine-tuned model, and the DeePMD reference values \cite{deangelis2025}.}
\begin{tabular}{lcccc}
\hline 
& $E_a$ & $D_o$ &  $D$ (\SI{450}{\kelvin})  &  $D$ (\SI{500}{\kelvin}) \\
& ($eV$)&  ($cm^2/s$) $\times 10^{-3}$ & ($cm^2/s$) $\times 10^{-6}$  & ($cm^2/s$) $\times 10^{-6}$\\ 
\hline
MACE-MPA-0 & $0.22 \pm 0.015$ & $1.1 \pm 0.42 $ & $3.45 \pm 0.50 $ & $7.3\pm 1.2 $ \\ 
Fine-tuned & $0.20 \pm 0.004$ & $1.1 \pm 0.13$ & $6.4 \pm 0.61 $ & $10.5 \pm 0.43$     \\ 
DeePMD & $0.24 $  & $3.3 $  &  $6.8$ &  $12.5 $ \\ 
\hline 
\end{tabular}
\label{tab:comparison}
\end{table*}
We find both foundational model and a model fine-tuned with only 300 data points to yield fair performance. 
To compare with DFT, Zheng {\it et al}~\cite{zheng2021lithium}, found the activation energy of 0.25 eV using DFT and the Climbing Image Nudged Elastic Band (CI-NEB) method on a 2×2×2 supercell. 
Similarly, Yildirim {\it et al}~\cite{First2015Yildirim}, reported a value of ~0.27 eV using the DFT and NEB method with the same supercell size.
It is worth noting that using larger supercells or including entropic effects is difficult with standard DFT.
Previous attempts of overcoming these limitations with ReaxFF potential, have not been successful despite extensive reparameterization effort, the structure failed to remain stable (liquefying during simulation) and yielded an activation energy of only 0.06 eV~\cite{deAngelis2024enhancing}.
Notably, the atoms involved in the knock-off event fluctuate in pairs (Fig.~\ref{fig:knockoff} (b)) and are typically preceded by multiple unsuccessful escape attempts
After knock-off, one atom displaces another, so while an individual atom moves a distance $x$, the net charge displacement is $2x$.
The Haven ratio (H$_r$) can provide insights into the effect of ionic correlations on diffusivity~\cite{Winter_2023, vargas2020dynamic}.
Computing H$_r$ remains an interesting direction and is left for future work.
Next, we investigate the impact of different fine-tuning strategies on model performance.

\subsection{Comparison of different fine-tuning strategies}
\begin{figure}[!ht]
\centering
\begin{subfigure}{0.49\textwidth} 
\includegraphics[width=\columnwidth]{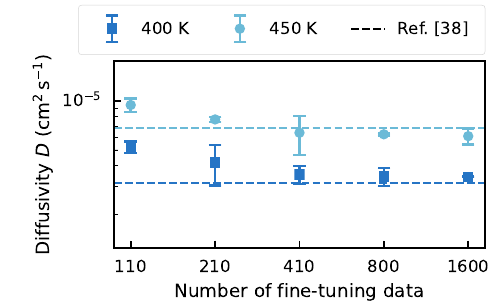}
  \caption{Fixed pre-training dataset of 1000 configuration and variable fine-tuning datasets between 110 to 800 data points.}
  \label{fig:fix_pt}
\end{subfigure}
\hfill
\begin{subfigure}{0.49\textwidth}
\includegraphics[width=\columnwidth]{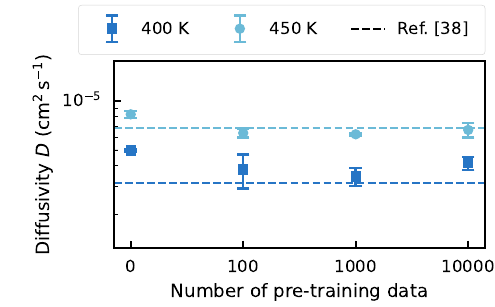}
  \caption{Fixed fine-tuning dataset of 800 data points: 700 for Li interstitial in LiF, and 100 for bulk LiF and variable pre-training dataset.}
  \label{fig:800}
\end{subfigure}
\hfill
\begin{subfigure}{0.49\textwidth}
\includegraphics[width=\columnwidth]{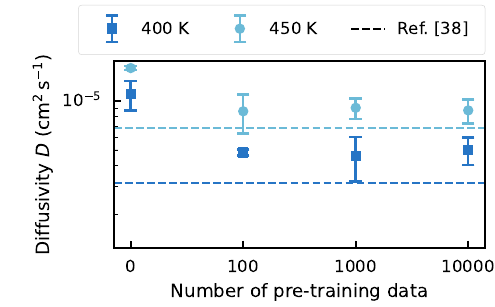}
\caption{Fixed fine-tuning dataset of 710 fine-tuning data points: 700 for Li interstitial in LiF, and 10 for bulk LiF and variable pre-training dataset.}
\centering
    \label{fig:710}
\end{subfigure}
\hfill
\begin{subfigure}{0.49\textwidth}
    \includegraphics[width=\columnwidth]{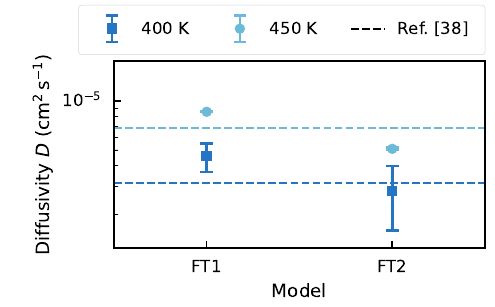}
    \caption{FT1 and FT2 models using configurations generated from MACE-MPA-0 trajectories (see Table.~\ref{tab:active_setup}) and no pre-training data.}
    \label{fig:my-data}
\end{subfigure}
\caption{The diffusivity at 400 K and 450 K computed using 3 ns trajectories from two independent replicas for different fine-tuned MACE models.
The models in (a), (b), and (c) are trained with DeePMD model generated dataset~\cite{deangelis2025}.
Note that the 800 data points in (b) contains the same 710 data points in (c) with 90 additional LiF bulk structures. 
The models in (d) were trained using data generated from configurations explored by the MACE-MPA-0 model.
The dashed lines are the DeePMD reference value~\cite{deangelis2025}.}
\label{fig:all}
\end{figure}

Following the fair performance of the MACE models observed in the previous section in predicting Arrhenius properties, we now consider the effect of different fine-tuning strategies. 
This comparison is designed to provide some insights into the training protocol.
The diffusivity of interstitial Li in LiF at 400 K and 450 K was found for several fine-tuned models using 3 ns trajectories in two independent replicas. 
Figs.~\ref{fig:fix_pt},~\ref{fig:800},~\ref{fig:710} show fine-tuned MACE-MPA-0 using the actively learned DeePMD data available in Ref.~\cite{deangelis2025}. Fig.~\ref{fig:my-data}, on the other hand, shows models trained using the configurations we generated from the trajectories explored by the MACE-MPA-0 potential (see Sections.~\ref{Fine-tuning with DeePMD data} and \ref{MPA-data}, respectively).

In Fig. \ref{fig:fix_pt}, the pre-training data set was fixed to 1000 randomly selected structures from MPtraj, the same data for all experiments, and the fine-tuning data was increasing from 110, 210, 410, 800 to 1600-samples datasets. 
All models exhibit good performance, with accuracy improving as the amount of fine-tuning data increases, gradually approaching the reference values.
At 800 data points, the model reaches good agreement with the reference and doubling the size of the data set provides no noticeable advantage.

On the other hand, in Fig. \ref{fig:800}, when the fine-tuning data set is fixed to 800, and the size of the pre-training dataset is varied from 0 and up to 10,000, we observe that increasing the amount of pre-training data initially improves agreement with the reference values. 
However, beyond a certain point, additional pre-training data yields diminishing returns.
A similar pattern can be observed with 200 fine-tuned data models although in this case the diminishing returns can be already observed with smaller pre-training dataset size that exceeds 100 data points (see Fig. S6 in Supplementary Information).
It is worth noting that in the 1600-point fine-tuning sets in Fig.~\ref{fig:fix_pt}, the pre-training dataset of 1000 structures continued to result in a good performance.
These results suggest that the smaller the fine-tuning dataset, the more sensitive the model is to the size of pre-training dataset.

Comparing Fig. \ref{fig:800} and Fig. \ref{fig:710}, the model trained on 710 data points (Fig.~\ref{fig:710}) overestimates diffusivity relative to the reference. This dataset consists of 700 interstitial Li in LiF configurations and only 10 bulk structures. When 90 additional bulk structures from the reference dataset are incorporated in the 800-point set (Fig.~\ref{fig:800}), the models prediction closely match the reference value.
Hence, achieving adequate representation of the underlying distribution is as important as the dataset size.

These results show that leveraging existing datasets (in this case, data generated through DeePMD active learning process), can provide the data for training accurate MACE models.
This observation is consistent with previous results of training MACE with datasets generated by finding the DFT energies and forces for configuration sampled from the trajectories explored by Gaussian Approximation Potential (GAP) models which yielded good performance~\cite{niblett2024transferability}.
In particular, when trained on GAP-generated data for conventional battery electrolyte solvents, it appears that MACE achieves better agreement with \emph{ab initio} diffusivity values compared to the original predictions from the GAP model \cite{niblett2024transferability}.
It is worth stressing that we found training a committee yields strong agreement between its members, indicating low uncertainty and thus eliminating the need for an active learning protocol.

Previous results used dataset carefully constructed for this specific problem via an active learning process with DeePMD potential~\cite{deangelis2025}, however, such resources are not always available.
As an alternative approach, we also investigate generating training data by sampling configurations from trajectories explored by the foundational model itself. 
Specifically, we sample configurations from a MACE-MPA-0 trajectory, compute their DFT energies, forces and stresses, and subsequently train a new fine-tuned MACE models.

Two models, FT1, trained on 156 structures: 16 bulk LiF, and 140 LiF with one interstitial Li atom, and model FT2, trained on 144 structures, 44 bulk LiF, and 100 LiF with one interstitial Li (see Table. \ref{tab:active_setup}). 
In other words, FT1 has less bulk configurations in its training data compared to the FT2.
Interestingly, in contrast to the behavior observed with DeePMD-generated data, we observe notable disagreement among committee members in the case of Li interestital in LiF, this necessitated using active learning protocol for this case (see Sec. \ref{sec:active_learning}).
The resulting lithium diffusivity values are shown in Fig. \ref{fig:my-data}. 
Both models demonstrate fair performance. 
FT1, which includes more interstitial configurations, slightly overestimates the diffusivity relative to the reference, while FT2, containing a larger proportion of bulk configurations, slightly underestimates it. 
Importantly, MACE achieves comparable accuracy to the DeePMD model \cite{deangelis2025}, despite being trained on merely 156 data points for FT1 and 144 data points for FT2 and does not require more than a single training round. 

It is also worth noting that the results obtained here surpass that obtained with ReaxFF, where the Radial Distribution Function (RDF) of the bulk is found to resemble a liquid phase rather than a solid phase and the activation energy is underestimated with a value of 0.06 eV even after long parametrization process that was highly sensitive to the training data~\cite{deAngelis2024enhancing}.

Model training was highly efficient and parallelizable, for example, fine-tuning the model with total 1800 data points, 800 fine-tuning data and 1000 pre-training data, with a batch size of 2 required just 4.5 hours to complete 800 epochs, averaging approximately 20 seconds per epoch (i.e., 20 seconds for going through all the data once) using 8 NVIDIA A100 GPUs.
Similarly, the model trained with similar settings but only 300 data points, 200 fine-tuning data and 100 pre-training data, needed an average of 6 seconds per epoch.
For running the MD simulations in LAMMPS, on a single NVIDIA A100 GPU, the simulation required approximately 0.24 seconds per step. With a time step of 1 fs and 300,000 steps this corresponds to 300 ps completed in 20 hours.
Multi-GPU support in LAMMPS has recently become available, efficient multi-GPU parallelization would significantly reduce the time reported here.
DeePMD, on the other hand, already have an efficient parallel implementation, a simulation of the same system can achieve nanoseconds per day. 
This highlights a trade-off between training data requirements and MD production speed.
For more complex systems, the need for additional training data grows, creating a bottleneck where DeePMD is constrained by the requirement for generating training data with computationally expensive DFT.

Constructing an optimal training dataset requires careful validation against the target property.
Some reports suggest that as few as 50 randomly selected data points~\cite{kovacs2023evaluation} are sufficient for training and as little as a single structure~\cite{deng2024overcoming} may suffice for fine-tuning foundational MLFF. 
While our results also demonstrate data efficiency and achieve fair results that is comparable to a well trained DeePMD model~\cite{deangelis2025} with substantially less data and effort. 
Nevertheless, we find that achieving good performance requires more data than these prior reports suggest
and ensuring accuracy and model robustness remains a non-trivial task.
As we showed, model performance depends on both the dataset size and its composition.
Since the training data directly shape the learned PES, assembling a broad-coverage and reliable dataset is essential, the amount of interstitial and bulk environments and the relative size of the pre-training dataset biases the model toward lower or higher mobility, and thus affects the diffusivity prediction.
%


\section{Conclusions}
In this study, we investigated interstitial Li diffusivity in LiF, a key SEI component that plays an important role in battery performance and safety.
Our results demonstrate that both the foundational MACE-MPA-0 model and its fine-tuned variants, trained with either data generated from configurations explored by a reference accurate model (DeePMD) or by the foundational MACE model itself, results in stable trajectories over long MD simulations and provide a good description of the system.
The use of the reference model data, however, offers a significant gain in training speed and ease of use.
Notably, the predicted values for activation energy and diffusivity are consistent with the values of the DeePMD model—trained on over 40,000 data points—despite MACE foundational model achieving comparable accuracy with little to no training.

In agreement with previous studies, we find that the data required for training MACE is minimal and that multi-head fine-tuning with pre-training data can improve model performance.
However, we observe that the choice of constituents and the ratio between different data groups (bulk, interstitial, pre-training) can bias the model, yielding either higher or lower diffusivity.
Furthermore, while the inclusion of pre-training data improves model performance, this benefit exhibits diminishing returns beyond a certain point that is dependent on the fine-tuning training dataset size. 
And while the benefits of pre-training dataset can be greater for smaller fine-tuning datasets, so too is the sensitivity to the pre-training dataset size.
These findings indicate that not only the size but also the proportionality between the constituents in the dataset needs to be accounted for and it confirms that rigorous model verification with respect to the specific property of interest remains critical.
Therefore, designing high-quality battery-specialized training datasets as well as developing better methods to determine the optimal structure of the training dataset can play a crucial role in rapidly adopting these computational tools to address critical, long-standing challenges in batteries.
That being said, the availability of such highly efficient models enables, for the first time, an in-depth analysis of the SEI formation mechanism, a process comprised of numerous components and a highly complex composition and topology, for which an efficient utilization of training data is highly needed in addition to accurate and computationally inexpensive models.

Beyond crystalline single materials of the dominant SEI components, an interesting direction is to investigate multi-component systems that covers heterogeneous and amorphous phases thereby more faithfully mimicking the complexity of real SEI.
In addition, although still computationally demanding, simulating the SEI formation process across different electrolytes and ion systems, including those beyond lithium-ion batteries, is becoming more feasible and would provide an important conjunction to experimental studies.
Finally, the good out-of-the-box performance of the foundational model can serve as an effective computational screening tool to identify highly diffusive materials for next-generation of battery technologies.

\section*{Acknowledgment}
This work was supported by computational resources provided by the CINECA IsB29 NEXT-LIB project and INRiM INR24-CHIROC project.

\section*{Funding}
E.C. acknowledges partial funding under the National Recovery and Resilience Plan 
(NRRP), (PNRR: Piano Nazionale di Ripresa e Resilienza), Mission 4 Component 2 Investment 1.3 - Call for tender No. 1561 dell'11.10.2022 del Ministero dell’Università e della Ricerca (MUR); funded by the European Union – NextGenerationE.

\section*{Roles}
CRediT: \textbf{NA}: Conceptualization, Data curation, Formal Analysis, Investigation, Methodology, Software, Validation, Visualization, Writing – original draft, Writing – review \& editing; \textbf{PDA}: Conceptualization, Project administration, Resources, Software, Supervision, Visualization, Writing – review \& editing; \textbf{PA} and \textbf{EC}: Conceptualization, Funding acquisition, Project administration, Resources, Supervision, Writing – review \& editing.

\section*{Data availability}
Data generated during this study will be made publicly available at \url{https://github.com/AlghamdiNada/Li_diffusion_with_MACE}

\section*{Supplementary Information}
Supplementary Information (13.6 MB PDF). The file includes supporting figures and tables.


\begin{thebibliography}{10}
\expandafter\ifx\csname url\endcsname\relax
  \def\url#1{\texttt{#1}}\fi
\expandafter\ifx\csname urlprefix\endcsname\relax\def\urlprefix{URL }\fi
\expandafter\ifx\csname href\endcsname\relax
  \def\href#1#2{#2} \def\path#1{#1}\fi

\bibitem{balakrishnan_great_2021}
N.~T.~M. Balakrishnan, A.~Das, N.~S. Jishnu, L.~R. Raphael, J.~D. Joyner, J.-H. Ahn, M.~J. Jabeen~Fatima, R.~Prasanth, {The Great History of {Lithium}-{Ion} {Batteries} and an {Overview} on {Energy} {Storage} {Devices}}, in: N.~T.~M. Balakrishnan, R.~Prasanth (Eds.), Electrospinning for {Advanced} {Energy} {Storage} {Applications}, Springer, Singapore, 2021, pp. 1--21.
\newblock \href {https://doi.org/10.1007/978-981-15-8844-0_1} {\path{doi:10.1007/978-981-15-8844-0_1}}.

\bibitem{amici_roadmap_2022}
J.~Amici, P.~Asinari, E.~Ayerbe, P.~Barboux, P.~Bayle‐Guillemaud, R.~J. Behm, M.~Berecibar, E.~Berg, A.~Bhowmik, S.~Bodoardo, I.~E. Castelli, I.~Cekic‐Laskovic, R.~Christensen, S.~Clark, R.~Diehm, R.~Dominko, M.~Fichtner, A.~A. Franco, A.~Grimaud, N.~Guillet, M.~Hahlin, S.~Hartmann, V.~Heiries, K.~Hermansson, A.~Heuer, S.~Jana, L.~Jabbour, J.~Kallo, A.~Latz, H.~Lorrmann, O.~M. Løvvik, S.~Lyonnard, M.~Meeus, E.~Paillard, S.~Perraud, T.~Placke, C.~Punckt, O.~Raccurt, J.~Ruhland, E.~Sheridan, H.~Stein, J.~Tarascon, V.~Trapp, T.~Vegge, M.~Weil, W.~Wenzel, M.~Winter, A.~Wolf, K.~Edström, A {Roadmap} for {Transforming} {Research} to {Invent} the {Batteries} of the {Future} {Designed} within the {European} {Large} {Scale} {Research} {Initiative} {BATTERY} 2030+, Advanced Energy Materials 12~(17) (2022) 2102785.
\newblock \href {https://doi.org/10.1002/aenm.202102785} {\path{doi:10.1002/aenm.202102785}}.

\bibitem{yoshio_lithium-ion_2009}
M.~Yoshio, R.~J. Brodd, A.~Kozawa (Eds.), Lithium-{Ion} {Batteries}: {Science} and {Technologies}, Springer, New York, NY, 2009.
\newblock \href {https://doi.org/10.1007/978-0-387-34445-4} {\path{doi:10.1007/978-0-387-34445-4}}.

\bibitem{goodenough_challenges_2010}
J.~B. Goodenough, Y.~Kim, Challenges for {Rechargeable} {Li} {Batteries}, Chemistry of Materials 22~(3) (2010) 587--603.
\newblock \href {https://doi.org/10.1021/cm901452z} {\path{doi:10.1021/cm901452z}}.

\bibitem{shan_brief_2021}
X.~Shan, Y.~Zhong, L.~Zhang, Y.~Zhang, X.~Xia, X.~Wang, J.~Tu, A {Brief} {Review} on {Solid} {Electrolyte} {Interphase} {Composition} {Characterization} {Technology} for {Lithium} {Metal} {Batteries}: {Challenges} and {Perspectives}, The Journal of Physical Chemistry C 125~(35) (2021) 19060--19080.
\newblock \href {https://doi.org/10.1021/acs.jpcc.1c06277} {\path{doi:10.1021/acs.jpcc.1c06277}}.

\bibitem{wang_fluorination_2024}
Y.~Wang, Z.~Wu, F.~M. Azad, Y.~Zhu, L.~Wang, C.~J. Hawker, A.~K. Whittaker, M.~Forsyth, C.~Zhang, Fluorination in advanced battery design, Nature Reviews Materials 9~(2) (2024) 119--133.
\newblock \href {https://doi.org/10.1038/s41578-023-00623-4} {\path{doi:10.1038/s41578-023-00623-4}}.

\bibitem{Palacin_2024}
M.~R. Palacin, P.~Johansson, R.~Dominko, B.~Dlugatch, D.~Aurbach, Z.~Li, M.~Fichtner, O.~Lužanin, J.~Bitenc, Z.~Wei, C.~Glaser, J.~Janek, A.~Fernández-Barquín, A.~R. Mainar, O.~Leonet, I.~Urdampilleta, J.~A. Blázquez, D.~S. Tchitchekova, A.~Ponrouch, P.~Canepa, G.~S. Gautam, R.~S. R.~G. Casilda, C.~S. Martinez-Cisneros, N.~U. Torres, A.~Varez, J.-Y. Sanchez, K.~V. Kravchyk, M.~V. Kovalenko, A.~A. Teck, H.~Shiel, I.~E.~L. Stephens, M.~P. Ryan, E.~Zemlyanushin, S.~Dsoke, R.~Grieco, N.~Patil, R.~Marcilla, X.~Gao, C.~J. Carmalt, G.~He, M.-M. Titirici, Roadmap on multivalent batteries, Journal of Physics: Energy 6~(3) (2024) 031501.
\newblock \href {https://doi.org/10.1088/2515-7655/ad34fc} {\path{doi:10.1088/2515-7655/ad34fc}}.

\bibitem{Titirici_2024}
M.~Titirici, P.~Johansson, M.~Crespo~Ribadeneyra, H.~Au, A.~Innocenti, S.~Passerini, E.~Petavratzi, P.~Lusty, A.~A. Tidblad, A.~J. Naylor, R.~Younesi, Y.~A. Chart, J.~Aspinall, M.~Pasta, J.~Orive, L.~M. Babulal, M.~Reynaud, K.~G. Latham, T.~Hosaka, S.~Komaba, J.~Bitenc, A.~Ponrouch, H.~Zhang, M.~Armand, R.~Kerr, P.~C. Howlett, M.~Forsyth, J.~Brown, A.~Grimaud, M.~Vilkman, K.~B. Dermenci, S.~Mousavihashemi, M.~Berecibar, J.~E. Marshall, C.~R. McElroy, E.~Kendrick, T.~Safdar, C.~Huang, F.~M. Zanotto, J.~F. Troncoso, D.~Z. Dominguez, M.~Alabdali, U.~Vijay, A.~A. Franco, S.~Pazhaniswamy, P.~S. Grant, S.~López~Guzman, M.~Fehse, M.~Galceran, N.~Antuñano, 2024 roadmap for sustainable batteries, Journal of Physics: Energy 6~(4) (2024) 041502.
\newblock \href {https://doi.org/10.1088/2515-7655/ad6bc0} {\path{doi:10.1088/2515-7655/ad6bc0}}.

\bibitem{cappabianca2023overview}
R.~Cappabianca, P.~De~Angelis, M.~Fasano, E.~Chiavazzo, P.~Asinari, An overview on transport phenomena within solid electrolyte interphase and their impact on the performance and durability of lithium-ion batteries, Energies 16~(13) (2023) 5003.
\newblock \href {https://doi.org/10.3390/en16135003} {\path{doi:10.3390/en16135003}}.

\bibitem{deAngelis2024enhancing}
P.~De~Angelis, R.~Cappabianca, M.~Fasano, P.~Asinari, E.~Chiavazzo, {Enhancing ReaxFF for molecular dynamics simulations of lithium-ion batteries: an interactive reparameterization protocol}, Scientific Reports 14~(1) (2024) 978.
\newblock \href {https://doi.org/10.1038/s41598-023-50978-5} {\path{doi:10.1038/s41598-023-50978-5}}.

\bibitem{Generalized2007Behler}
J.~Behler, M.~Parrinello, {Generalized Neural-Network Representation of High-Dimensional Potential-Energy Surfaces}, Phys. Rev. Lett. 98 (2007) 146401.
\newblock \href {https://doi.org/10.1103/PhysRevLett.98.146401} {\path{doi:10.1103/PhysRevLett.98.146401}}.

\bibitem{Unke2021Machine}
O.~T. Unke, S.~Chmiela, H.~E. Sauceda, M.~Gastegger, I.~Poltavsky, K.~T. Sch{\"u}tt, A.~Tkatchenko, K.-R. M{\"u}ller, {Machine Learning Force Fields}, Chemical Reviews 121~(16) (2021) 10142--10186, pMID: 33705118.
\newblock \href {https://doi.org/10.1021/acs.chemrev.0c01111} {\path{doi:10.1021/acs.chemrev.0c01111}}.

\bibitem{Deringer_2020}
V.~L. Deringer, Modelling and understanding battery materials with machine-learning-driven atomistic simulations, Journal of Physics: Energy 2~(4) (2020) 041003.
\newblock \href {https://doi.org/10.1088/2515-7655/abb011} {\path{doi:10.1088/2515-7655/abb011}}.

\bibitem{CASINI2024100330}
M.~Casini, P.~De~Angelis, E.~Chiavazzo, L.~Bergamasco, Current trends on the use of deep learning methods for image analysis in energy applications, Energy and AI 15 (2024) 100330.
\newblock \href {https://doi.org/10.1016/j.egyai.2023.100330} {\path{doi:10.1016/j.egyai.2023.100330}}.

\bibitem{battery2030}
M.~Fichtner, M.~Fichtner, K.~Edström, E.~Ayerbe, M.~Berecibar, A.~Bhowmik, I.~E. Castelli, S.~Clark, R.~Dominko, M.~Erakca, A.~A. Franco, A.~Grimaud, B.~Horstmann, A.~Latz, H.~Lorrmann, M.~Meeus, R.~Narayan, F.~Pammer, J.~Ruhland, H.~Stein, T.~Vegge, M.~Weil, {Rechargeable Batteries of the Future—The State of the Art from a BATTERY 2030+ Perspective}, Advanced Energy Materials 12~(17) (2022) 2102904.
\newblock \href {https://doi.org/10.1002/aenm.202102904} {\path{doi:10.1002/aenm.202102904}}.

\bibitem{Zhang2018Deep}
L.~Zhang, J.~Han, H.~Wang, R.~Car, W.~E, Deep potential molecular dynamics: A scalable model with the accuracy of quantum mechanics, Phys. Rev. Lett. 120 (2018) 143001.
\newblock \href {https://doi.org/10.1103/PhysRevLett.120.143001} {\path{doi:10.1103/PhysRevLett.120.143001}}.

\bibitem{deepmd}
J.~Han, L.~Zhang, R.~Car, W.~E, {Deep potential: A general representation of a many-body potential energy surface}, Communications in Computational Physics (2018) 629–639\href {https://doi.org/10.4208/cicp.oa-2017-0213} {\path{doi:10.4208/cicp.oa-2017-0213}}.

\bibitem{Zeng2023DeePMD}
J.~Zeng, D.~Zhang, D.~Lu, P.~Mo, Z.~Li, Y.~Chen, M.~Rynik, L.~Huang, Z.~Li, S.~Shi, Y.~Wang, H.~Ye, P.~Tuo, J.~Yang, Y.~Ding, Y.~Li, D.~Tisi, Q.~Zeng, H.~Bao, Y.~Xia, J.~Huang, K.~Muraoka, Y.~Wang, J.~Chang, F.~Yuan, S.~L. Bore, C.~Cai, Y.~Lin, B.~Wang, J.~Xu, J.-X. Zhu, C.~Luo, Y.~Zhang, R.~E.~A. Goodall, W.~Liang, A.~K. Singh, S.~Yao, J.~Zhang, R.~Wentzcovitch, J.~Han, J.~Liu, W.~Jia, D.~M. York, W.~E, R.~Car, L.~Zhang, H.~Wang, {DeePMD-kit v2: A software package for deep potential models}, The Journal of Chemical Physics 159~(5) (2023) 054801.
\newblock \href {https://doi.org/10.1063/5.0155600} {\path{doi:10.1063/5.0155600}}.

\bibitem{Wang_2018}
H.~Wang, L.~Zhang, J.~Han, W.~E, {DeePMD-kit: A deep learning package for many-body potential energy representation and molecular dynamics}, Computer Physics Communications 228 (2018) 178–184.
\newblock \href {https://doi.org/10.1016/j.cpc.2018.03.016} {\path{doi:10.1016/j.cpc.2018.03.016}}.

\bibitem{Thiemann_2025}
F.~L. Thiemann, N.~O’Neill, V.~Kapil, A.~Michaelides, C.~Schran, Introduction to machine learning potentials for atomistic simulations, Journal of Physics: Condensed Matter 37~(7) (2024) 073002.
\newblock \href {https://doi.org/10.1088/1361-648X/ad9657} {\path{doi:10.1088/1361-648X/ad9657}}.

\bibitem{gilmer2017neuralmessagepassingquantum}
J.~Gilmer, S.~S. Schoenholz, P.~F. Riley, O.~Vinyals, G.~E. Dahl, {Neural Message Passing for Quantum Chemistry} (2017).
\newblock \href {https://doi.org/10.48550/arXiv.1704.01212} {\path{doi:10.48550/arXiv.1704.01212}}.

\bibitem{batatia2023macehigherorderequivariant}
I.~Batatia, D.~P. Kovács, G.~N.~C. Simm, C.~Ortner, G.~Csányi, {MACE: Higher Order Equivariant Message Passing Neural Networks for Fast and Accurate Force Fields} (2023).
\newblock \href {https://doi.org/10.48550/arXiv.2206.07697} {\path{doi:10.48550/arXiv.2206.07697}}.

\bibitem{ACE}
R.~Drautz, Atomic cluster expansion for accurate and transferable interatomic potentials, Phys. Rev. B 99 (2019) 014104.
\newblock \href {https://doi.org/10.1103/PhysRevB.99.014104} {\path{doi:10.1103/PhysRevB.99.014104}}.

\bibitem{batatia2024foundationmodelatomisticmaterials}
I.~Batatia, P.~Benner, Y.~Chiang, A.~M. Elena, D.~P. Kovács, J.~Riebesell, X.~R. Advincula, M.~Asta, M.~Avaylon, W.~J. Baldwin, F.~Berger, N.~Bernstein, A.~Bhowmik, S.~M. Blau, V.~Cărare, J.~P. Darby, S.~De, F.~D. Pia, V.~L. Deringer, R.~Elijošius, Z.~El-Machachi, F.~Falcioni, E.~Fako, A.~C. Ferrari, A.~Genreith-Schriever, J.~George, R.~E.~A. Goodall, C.~P. Grey, P.~Grigorev, S.~Han, W.~Handley, H.~H. Heenen, K.~Hermansson, C.~Holm, J.~Jaafar, S.~Hofmann, K.~S. Jakob, H.~Jung, V.~Kapil, A.~D. Kaplan, N.~Karimitari, J.~R. Kermode, N.~Kroupa, J.~Kullgren, M.~C. Kuner, D.~Kuryla, G.~Liepuoniute, J.~T. Margraf, I.-B. Magdău, A.~Michaelides, J.~H. Moore, A.~A. Naik, S.~P. Niblett, S.~W. Norwood, N.~O'Neill, C.~Ortner, K.~A. Persson, K.~Reuter, A.~S. Rosen, L.~L. Schaaf, C.~Schran, B.~X. Shi, E.~Sivonxay, T.~K. Stenczel, V.~Svahn, C.~Sutton, T.~D. Swinburne, J.~Tilly, C.~van~der Oord, E.~Varga-Umbrich, T.~Vegge, M.~Vondrák, Y.~Wang, W.~C. Witt, F.~Zills, G.~Csányi, A foundation model for atomistic materials
  chemistry (2024).
\newblock \href {https://doi.org/10.48550/arXiv.2401.00096} {\path{doi:10.48550/arXiv.2401.00096}}.

\bibitem{Haochen2024systematic}
H.~Yu, M.~Giantomassi, G.~Materzanini, J.~Wang, G.-M. Rignanese, Systematic assessment of various universal machine-learning interatomic potentials, Materials Genome Engineering Advances 2~(3) (2024) e58.
\newblock \href {https://doi.org/10.1002/mgea.58} {\path{doi:10.1002/mgea.58}}.

\bibitem{Bruno2025performance}
B.~Focassio, L.~P. M.~Freitas, G.~R. Schleder, {Performance Assessment of Universal Machine Learning Interatomic Potentials: Challenges and Directions for Materials’ Surfaces}, ACS Applied Materials \& Interfaces 17~(9) (2025) 13111--13121, pMID: 38990833.
\newblock \href {https://doi.org/10.1021/acsami.4c03815} {\path{doi:10.1021/acsami.4c03815}}.

\bibitem{Deng2023CHGNet}
B.~Deng, P.~Zhong, K.~Jun, J.~Riebesell, K.~Han, C.~J. Bartel, G.~Ceder, {CHGNet as a pretrained universal neural network potential for charge-informed atomistic modelling}, Nature Machine Intelligence 5 (2023) 1031--1041.
\newblock \href {https://doi.org/10.1038/s42256-023-00716-3} {\path{doi:10.1038/s42256-023-00716-3}}.

\bibitem{Alex}
J.~Schmidt, T.~F. Cerqueira, A.~H. Romero, A.~Loew, F.~Jäger, H.-C. Wang, S.~Botti, M.~A. Marques, Improving machine-learning models in materials science through large datasets, Materials Today Physics 48 (2024) 101560.
\newblock \href {https://doi.org/10.1016/j.mtphys.2024.101560} {\path{doi:10.1016/j.mtphys.2024.101560}}.

\bibitem{barrosoluque2024openmaterials2024omat24}
L.~Barroso-Luque, M.~Shuaibi, X.~Fu, B.~M. Wood, M.~Dzamba, M.~Gao, A.~Rizvi, C.~L. Zitnick, Z.~W. Ulissi, {Open Materials 2024 (OMat24) Inorganic Materials Dataset and Models} (2024).
\newblock \href {https://doi.org/10.48550/arXiv.2410.12771} {\path{doi:10.48550/arXiv.2410.12771}}.

\bibitem{Church_Chen_Ma_2021}
K.~W. Church, Z.~Chen, Y.~Ma, {Emerging trends: A gentle introduction to fine-tuning}, Natural Language Engineering 27~(6) (2021) 763–778.
\newblock \href {https://doi.org/10.1017/S1351324921000322} {\path{doi:10.1017/S1351324921000322}}.

\bibitem{D4FD00107A}
H.~Kaur, F.~Della~Pia, I.~Batatia, X.~R. Advincula, B.~X. Shi, J.~Lan, G.~Csányi, A.~Michaelides, V.~Kapil, Data-efficient fine-tuning of foundational models for first-principles quality sublimation enthalpies, Faraday Discuss. 256 (2025) 120--138.
\newblock \href {https://doi.org/10.1039/D4FD00107A} {\path{doi:10.1039/D4FD00107A}}.

\bibitem{BHOWMIK2019446}
A.~Bhowmik, I.~E. Castelli, J.~M. Garcia-Lastra, P.~B. Jørgensen, O.~Winther, T.~Vegge, A perspective on inverse design of battery interphases using multi-scale modelling, experiments and generative deep learning, Energy Storage Materials 21 (2019) 446--456.
\newblock \href {https://doi.org/https://doi.org/10.1016/j.ensm.2019.06.011} {\path{doi:https://doi.org/10.1016/j.ensm.2019.06.011}}.

\bibitem{WU2021106489}
J.~Wu, M.~Ihsan-Ul-Haq, Y.~Chen, J.-K. Kim, Understanding solid electrolyte interphases: Advanced characterization techniques and theoretical simulations, Nano Energy 89 (2021) 106489.
\newblock \href {https://doi.org/https://doi.org/10.1016/j.nanoen.2021.106489} {\path{doi:https://doi.org/10.1016/j.nanoen.2021.106489}}.

\bibitem{Zhang_2023}
C.~Zhang, J.~Cheng, Y.~Chen, M.~K.~Y. Chan, Q.~Cai, R.~P. Carvalho, C.~F.~N. Marchiori, D.~Brandell, C.~M. Araujo, M.~Chen, X.~Ji, G.~Feng, K.~Goloviznina, A.~Serva, M.~Salanne, T.~Mandai, T.~Hosaka, M.~Alhanash, P.~Johansson, Y.-Z. Qiu, H.~Xiao, M.~Eikerling, R.~Jinnouchi, M.~M. Melander, G.~Kastlunger, A.~Bouzid, A.~Pasquarello, S.-J. Shin, M.~M. Kim, H.~Kim, K.~Schwarz, R.~Sundararaman, 2023 roadmap on molecular modelling of electrochemical energy materials, Journal of Physics: Energy 5~(4) (2023) 041501.
\newblock \href {https://doi.org/10.1088/2515-7655/acfe9b} {\path{doi:10.1088/2515-7655/acfe9b}}.

\bibitem{senftle2016reaxff}
T.~P. Senftle, S.~Hong, M.~M. Islam, S.~B. Kylasa, Y.~Zheng, Y.~K. Shin, C.~Junkermeier, R.~Engel-Herbert, M.~J. Janik, H.~M. Aktulga, et~al., {The ReaxFF reactive force-field: development, applications and future directions}, npj Computational Materials 2~(1) (2016) 15011.
\newblock \href {https://doi.org/10.1038/npjcompumats.2015.11} {\path{doi:10.1038/npjcompumats.2015.11}}.

\bibitem{vanDuin2001ReaxFF}
A.~C.~T. van Duin, S.~Dasgupta, F.~Lorant, W.~A. Goddard, {ReaxFF: A Reactive Force Field for Hydrocarbons}, The Journal of Physical Chemistry A 105~(41) (2001) 9396--9409.
\newblock \href {https://doi.org/10.1021/jp004368u} {\path{doi:10.1021/jp004368u}}.

\bibitem{li2025enabling}
W.-Q. Li, G.~Wu, J.~M. Arce-Ramos, Y.~H. Lau, M.-F. Ng, Enabling accurate modelling of materials for a solid electrolyte interphase in lithium-ion batteries using effective machine learning interatomic potentials, Materials Horizons 12~(24) (2025) 10770--10781.
\newblock \href {https://doi.org/10.1039/D5MH01343G} {\path{doi:10.1039/D5MH01343G}}.

\bibitem{deangelis2025}
P.~De~Angelis, U.~Raucci, F.~Mambretti, M.~Fasano, E.~Chiavazzo, P.~Asinari, M.~Parrinello, {Exploring Lithium Diffusion in LiF with Machine Learning Potentials: From Point Defects to Collective Ring Diffusion}, ChemRxiv (2025).
\newblock \href {https://doi.org/10.26434/chemrxiv-2025-fv0zl} {\path{doi:10.26434/chemrxiv-2025-fv0zl}}.

\bibitem{kovacs2023evaluation}
D.~P. Kov{\'a}cs, I.~Batatia, E.~S. Arany, G.~Cs{\'a}nyi, Evaluation of the mace force field architecture: From medicinal chemistry to materials science, The Journal of Chemical Physics 159~(4) (2023) 044118.
\newblock \href {https://doi.org/10.1063/5.0155322} {\path{doi:10.1063/5.0155322}}.

\bibitem{fu2023forcesenoughbenchmarkcritical}
X.~Fu, Z.~Wu, W.~Wang, T.~Xie, S.~Keten, R.~Gomez-Bombarelli, T.~Jaakkola, Forces are not enough: Benchmark and critical evaluation for machine learning force fields with molecular simulations (2023).
\newblock \href {https://doi.org/10.48550/arXiv.2210.07237} {\path{doi:10.48550/arXiv.2210.07237}}.

\bibitem{riebesell2025framework}
J.~Riebesell, R.~E. Goodall, P.~Benner, Y.~Chiang, B.~Deng, G.~Ceder, M.~Asta, A.~A. Lee, A.~Jain, K.~A. Persson, A framework to evaluate machine learning crystal stability predictions, nature machine intelligence 7~(6) (2025) 836--847.
\newblock \href {https://doi.org/10.1038/s42256-025-01055-1} {\path{doi:10.1038/s42256-025-01055-1}}.

\bibitem{Yu2024Systematic}
H.~Yu, M.~Giantomassi, G.~Materzanini, J.~Wang, G.-M. Rignanese, Systematic assessment of various universal machine-learning interatomic potentials, Materials Genome Engineering Advances 2~(3) (2024) e58.
\newblock \href {https://doi.org/https://doi.org/10.1002/mgea.58} {\path{doi:https://doi.org/10.1002/mgea.58}}.

\bibitem{Wines2025CHIPS}
D.~Wines, K.~Choudhary, Chips-ff: Evaluating universal machine learning force fields for material properties, ACS Materials Letters 7~(6) (2025) 2105--2114.
\newblock \href {https://doi.org/10.1021/acsmaterialslett.5c00093} {\path{doi:10.1021/acsmaterialslett.5c00093}}.

\bibitem{Li2025Are}
D.~Li, J.~Yang, X.~Chen, L.~Yu, S.~Liu, Are foundational atomistic models reliable for finite-temperature molecular dynamics?, The Journal of Physical Chemistry C 129~(49) (2025) 21538--21544.
\newblock \href {https://doi.org/10.1021/acs.jpcc.5c07541} {\path{doi:10.1021/acs.jpcc.5c07541}}.

\bibitem{chiang2025mliparenaadvancingfairness}
Y.~Chiang, T.~Kreiman, C.~Zhang, M.~C. Kuner, E.~Weaver, I.~Amin, H.~Park, Y.~Lim, J.~Kim, D.~Chrzan, A.~Walsh, S.~M. Blau, M.~Asta, A.~S. Krishnapriyan, Mlip arena: Advancing fairness and transparency in machine learning interatomic potentials via an open, accessible benchmark platform (2025).
\newblock \href {https://doi.org/10.48550/arXiv.2509.20630} {\path{doi:10.48550/arXiv.2509.20630}}.

\bibitem{LAMMPS}
A.~P. Thompson, H.~M. Aktulga, R.~Berger, D.~S. Bolintineanu, W.~M. Brown, P.~S. Crozier, P.~J. in~'t Veld, A.~Kohlmeyer, S.~G. Moore, T.~D. Nguyen, R.~Shan, M.~J. Stevens, J.~Tranchida, C.~Trott, S.~J. Plimpton, {LAMMPS} - a flexible simulation tool for particle-based materials modeling at the atomic, meso, and continuum scales, Comp. Phys. Comm. 271 (2022) 108171.
\newblock \href {https://doi.org/10.1016/j.cpc.2021.108171} {\path{doi:10.1016/j.cpc.2021.108171}}.

\bibitem{ase-paper}
A.~H. Larsen, J.~J. Mortensen, J.~Blomqvist, I.~E. Castelli, R.~Christensen, M.~Dułak, J.~Friis, M.~N. Groves, B.~Hammer, C.~Hargus, E.~D. Hermes, P.~C. Jennings, P.~B. Jensen, J.~Kermode, J.~R. Kitchin, E.~L. Kolsbjerg, J.~Kubal, K.~Kaasbjerg, S.~Lysgaard, J.~B. Maronsson, T.~Maxson, T.~Olsen, L.~Pastewka, A.~Peterson, C.~Rostgaard, J.~Schiøtz, O.~Schütt, M.~Strange, K.~S. Thygesen, T.~Vegge, L.~Vilhelmsen, M.~Walter, Z.~Zeng, K.~W. Jacobsen, {The atomic simulation environment—a Python library for working with atoms}, Journal of Physics: Condensed Matter 29~(27) (2017) 273002.
\newblock \href {https://doi.org/10.1088/1361-648X/aa680e} {\path{doi:10.1088/1361-648X/aa680e}}.

\bibitem{Batatia2022Design}
I.~Batatia, S.~Batzner, D.~P. Kov{\'a}cs, A.~Musaelian, G.~N.~C. Simm, R.~Drautz, C.~Ortner, B.~Kozinsky, G.~Cs{\'a}nyi, {The Design Space of E(3)-Equivariant Atom-Centered Interatomic Potentials} (2022).
\newblock \href {https://doi.org/10.48550/arXiv.2205.06643} {\path{doi:10.48550/arXiv.2205.06643}}.

\bibitem{nose1984unified}
S.~Nos{\'e}, {A unified formulation of the constant temperature molecular dynamics methods}, The Journal of chemical physics 81~(1) (1984) 511--519.
\newblock \href {https://doi.org/10.1063/1.447334} {\path{doi:10.1063/1.447334}}.

\bibitem{hoover1985canonical}
W.~G. Hoover, {Canonical dynamics: Equilibrium phase-space distributions}, Physical review A 31~(3) (1985) 1695.
\newblock \href {https://doi.org/10.1103/PhysRevA.31.1695} {\path{doi:10.1103/PhysRevA.31.1695}}.

\bibitem{allen2017computer}
M.~P. Allen, D.~J. Tildesley, Computer simulation of liquids, Oxford university press, 2017.

\bibitem{Mehrer2007}
H.~Mehrer, Diffusion in Solids, Springer Berlin Heidelberg, Berlin, Heidelberg, 2007, Ch. Dependence of Diffusion on Temperature and Pressure, pp. 127--149.
\newblock \href {https://doi.org/10.1007/978-3-540-71488-0_8} {\path{doi:10.1007/978-3-540-71488-0_8}}.

\bibitem{mace_mpa_0}
I.~Batatia, P.~Benner, Y.~Chiang, A.~M. Elena, D.~P. Kovács, J.~Riebesell, X.~R. Advincula, M.~Asta, W.~J. Baldwin, N.~Bernstein, A.~Bhowmik, S.~M. Blau, V.~Cărare, J.~P. Darby, S.~De, F.~D. Pia, V.~L. Deringer, R.~Elijošius, Z.~El-Machachi, E.~Fako, A.~C. Ferrari, A.~Genreith-Schriever, J.~George, R.~E.~A. Goodall, C.~P. Grey, S.~Han, W.~Handley, H.~H. Heenen, K.~Hermansson, C.~Holm, J.~Jaafar, S.~Hofmann, K.~S. Jakob, H.~Jung, V.~Kapil, A.~D. Kaplan, N.~Karimitari, N.~Kroupa, J.~Kullgren, M.~C. Kuner, D.~Kuryla, G.~Liepuoniute, J.~T. Margraf, I.-B. Magdău, A.~Michaelides, J.~H. Moore, A.~A. Naik, S.~P. Niblett, S.~W. Norwood, N.~O'Neill, C.~Ortner, K.~A. Persson, K.~Reuter, A.~S. Rosen, L.~L. Schaaf, C.~Schran, E.~Sivonxay, T.~K. Stenczel, V.~Svahn, C.~Sutton, C.~van~der Oord, E.~Varga-Umbrich, T.~Vegge, M.~Vondrák, Y.~Wang, W.~C. Witt, F.~Zills, G.~Csányi, {MACE-MPA-0}, \url{https://github.com/ACEsuit/mace-foundations/releases/tag/mace_mpa_0} (2024).

\bibitem{Huber1992}
P.~J. Huber, {Robust Estimation of a Location Parameter}, Springer New York, New York, NY, 1992.
\newblock \href {https://doi.org/10.1007/978-1-4612-4380-9_35} {\path{doi:10.1007/978-1-4612-4380-9_35}}.

\bibitem{yang2023reactant}
M.~Yang, U.~Raucci, M.~Parrinello, Reactant-induced dynamics of lithium imide surfaces during the ammonia decomposition process, Nature Catalysis 6~(9) (2023) 829--836.
\newblock \href {https://doi.org/10.1038/s41929-023-01006-2} {\path{doi:10.1038/s41929-023-01006-2}}.

\bibitem{giannozzi2009quantum}
P.~Giannozzi, S.~Baroni, N.~Bonini, M.~Calandra, R.~Car, C.~Cavazzoni, D.~Ceresoli, G.~L. Chiarotti, M.~Cococcioni, I.~Dabo, A.~Dal~Corso, S.~de~Gironcoli, S.~Fabris, G.~Fratesi, R.~Gebauer, U.~Gerstmann, C.~Gougoussis, A.~Kokalj, M.~Lazzeri, L.~Martin-Samos, N.~Marzari, F.~Mauri, R.~Mazzarello, S.~Paolini, A.~Pasquarello, L.~Paulatto, C.~Sbraccia, S.~Scandolo, G.~Sclauzero, A.~P. Seitsonen, A.~Smogunov, P.~Umari, R.~M. Wentzcovitch, {QUANTUM ESPRESSO: a modular and open-source software project for quantumsimulations of materials}, Journal of physics: Condensed matter 21~(39) (2009) 395502.
\newblock \href {https://doi.org/10.1088/0953-8984/21/39/395502} {\path{doi:10.1088/0953-8984/21/39/395502}}.

\bibitem{Giannozzi_2017}
P.~Giannozzi, O.~Andreussi, T.~Brumme, O.~Bunau, M.~Buongiorno~Nardelli, M.~Calandra, R.~Car, C.~Cavazzoni, D.~Ceresoli, M.~Cococcioni, N.~Colonna, I.~Carnimeo, A.~Dal~Corso, S.~de~Gironcoli, P.~Delugas, R.~A. DiStasio, A.~Ferretti, A.~Floris, G.~Fratesi, G.~Fugallo, R.~Gebauer, U.~Gerstmann, F.~Giustino, T.~Gorni, J.~Jia, M.~Kawamura, H.-Y. Ko, A.~Kokalj, E.~Küçükbenli, M.~Lazzeri, M.~Marsili, N.~Marzari, F.~Mauri, N.~L. Nguyen, H.-V. Nguyen, A.~Otero-de-la Roza, L.~Paulatto, S.~Poncé, D.~Rocca, R.~Sabatini, B.~Santra, M.~Schlipf, A.~P. Seitsonen, A.~Smogunov, I.~Timrov, T.~Thonhauser, P.~Umari, N.~Vast, X.~Wu, S.~Baroni, Advanced capabilities for materials modelling with quantum espresso, Journal of Physics: Condensed Matter 29~(46) (2017) 465901.
\newblock \href {https://doi.org/10.1088/1361-648X/aa8f79} {\path{doi:10.1088/1361-648X/aa8f79}}.

\bibitem{Giannozzi2020exascale}
P.~Giannozzi, O.~Baseggio, P.~Bonfà, D.~Brunato, R.~Car, I.~Carnimeo, C.~Cavazzoni, S.~de~Gironcoli, P.~Delugas, F.~Ferrari~Ruffino, A.~Ferretti, N.~Marzari, I.~Timrov, A.~Urru, S.~Baroni, {Quantum ESPRESSO toward the exascale}, The Journal of Chemical Physics 152~(15) (2020) 154105.
\newblock \href {https://doi.org/10.1063/5.0005082} {\path{doi:10.1063/5.0005082}}.

\bibitem{PhysRevLett.77.3865}
J.~P. Perdew, K.~Burke, M.~Ernzerhof, {Generalized Gradient Approximation Made Simple}, Phys. Rev. Lett. 77 (1996) 3865--3868.
\newblock \href {https://doi.org/10.1103/PhysRevLett.77.3865} {\path{doi:10.1103/PhysRevLett.77.3865}}.

\bibitem{Tan2021Growing}
J.~Tan, J.~Matz, P.~Dong, J.~Shen, M.~Ye, A growing appreciation for the role of lif in the solid electrolyte interphase, Advanced Energy Materials 11~(16) (2021) 2100046.
\newblock \href {https://doi.org/https://doi.org/10.1002/aenm.202100046} {\path{doi:https://doi.org/10.1002/aenm.202100046}}.

\bibitem{zheng2021lithium}
J.~Zheng, Z.~Ju, B.~Zhang, J.~Nai, T.~Liu, Y.~Liu, Q.~Xie, W.~Zhang, Y.~Wang, X.~Tao, Lithium ion diffusion mechanism on the inorganic components of the solid--electrolyte interphase, Journal of Materials Chemistry A 9~(16) (2021) 10251--10259.
\newblock \href {https://doi.org/10.1039/D0TA11444H} {\path{doi:10.1039/D0TA11444H}}.

\bibitem{he2018statistical}
X.~He, Y.~Zhu, A.~Epstein, Y.~Mo, Statistical variances of diffusional properties from ab initio molecular dynamics simulations, npj Computational Materials 4~(1) (2018) 18.
\newblock \href {https://doi.org/10.1038/s41524-018-0074-y} {\path{doi:10.1038/s41524-018-0074-y}}.

\bibitem{deng2024overcoming}
B.~Deng, Y.~Choi, P.~Zhong, J.~Riebesell, S.~Anand, Z.~Li, K.~Jun, K.~A. Persson, G.~Ceder, {Overcoming systematic softening in universal machine learning interatomic potentials by fine-tuning} (2024).
\newblock \href {https://doi.org/10.48550/arXiv.2405.07105} {\path{doi:10.48550/arXiv.2405.07105}}.

\bibitem{First2015Yildirim}
H.~Yildirim, A.~Kinaci, M.~K.~Y. Chan, J.~P. Greeley, First-principles analysis of defect thermodynamics and ion transport in inorganic sei compounds: Lif and naf, ACS Applied Materials \& Interfaces 7~(34) (2015) 18985--18996, pMID: 26255641.
\newblock \href {https://doi.org/10.1021/acsami.5b02904} {\path{doi:10.1021/acsami.5b02904}}.

\bibitem{Winter_2023}
G.~Winter, R.~Gómez-Bombarelli, Simulations with machine learning potentials identify the ion conduction mechanism mediating non-arrhenius behavior in lgps, Journal of Physics: Energy 5~(2) (2023) 024004.
\newblock \href {https://doi.org/10.1088/2515-7655/acbbef} {\path{doi:10.1088/2515-7655/acbbef}}.

\bibitem{vargas2020dynamic}
N.~M. Vargas-Barbosa, B.~Roling, Dynamic ion correlations in solid and liquid electrolytes: how do they affect charge and mass transport?, ChemElectroChem 7~(2) (2020) 367--385.
\newblock \href {https://doi.org/10.1002/celc.201901627} {\path{doi:10.1002/celc.201901627}}.

\bibitem{niblett2024transferability}
S.~P. Niblett, P.~Kourtis, I.-B. Magdău, C.~P. Grey, G.~Csányi, Transferability of data sets between machine-learned interatomic potential algorithms, Journal of Chemical Theory and Computation 21~(12) (2025) 6096--6112, pMID: 40470788.
\newblock \href {https://doi.org/10.1021/acs.jctc.5c00272} {\path{doi:10.1021/acs.jctc.5c00272}}.

\end{thebibliography}
\end{document}